\documentclass{PoS}
\usepackage{color}
\usepackage[square, numbers, comma, sort&compress]{natbib}  % Use the "Natbib" style for the references in the Bibliography
\usepackage{verbatim}  % Needed for the "comment" environment to make LaTeX comments
\usepackage{vector}  % Allows "\bvec{}" and "\buvec{}" for "blackboard" style bold vectors in maths
\usepackage{amsmath}
\usepackage{amsfonts}
\usepackage{amssymb}
\usepackage{braket}
\usepackage{tabularx}
\usepackage{amsthm}
\usepackage{epsf}
\usepackage{booktabs}
\usepackage{subfigure}
\usepackage[T1]{fontenc}                
\usepackage{float}
\usepackage{graphicx}
\usepackage[section]{placeins}
\usepackage{natbib}

\title{$q\bar q$-potential: a numerical study}

\ShortTitle{$q\bar q$-potential simulations}

\author{ \speaker{A. Guerrieri} \\
     {Dipartimento di Fisica, Universit\`a di Roma {\it La Sapienza}\\ \small and INFN, Sezione di Roma 1 \\ \small P.le A.\ Moro 5, 00185 Roma, Italy}\\
      E-mail: \email{andrea.guerrieri@roma1.infn.it}}

\author{S. Petrarca\\
       {Dipartimento di Fisica, Universit\`a di Roma {\it La Sapienza}\\ \small and INFN, Sezione di Roma 1 \\ \small P.le A.\ Moro 5, 00185 Roma, Italy}\\
      E-mail: \email{silvano.petrarca@roma1.infn.it}}
        
        \author{A. Rubeo\\
     {Dipartimento di Fisica, Universit\`a di Roma {\it La Sapienza}\\ \small and INFN, Sezione di Roma 1 \\ \small P.le A.\ Moro 5, 00185 Roma, Italy}\\
      E-mail: \email{argia.rubeo@roma1.infn.it}}

\author{M. Testa\\
       {Dipartimento di Fisica, Universit\`a di Roma {\it La Sapienza}\\ \small and INFN, Sezione di Roma 1 \\ \small P.le A.\ Moro 5, 00185 Roma, Italy}
\\
E-mail: \email{massimo.testa@roma1.infn.it}      }

\abstract{We report the results of recent lattice simulations aimed at computing the $q$ and $\bar q$ potential energies in the singlet and the octet (adjoint) representation.}

\FullConference{31st International Symposium on Lattice Field Theory - LATTICE 2013\\
		July 29 - August 3, 2013\\
		Mainz, Germany}

\begin{document}

\section{Introduction}
\label{sec:INTRO}

It is shown in refs.~\citep{Rossi:2013qba} and~\citep{PROC} that in the temporal gauge formulation of the Yang--Mills theories introduced in refs.~\citep{Rossi:1979jf}~\citep{Rossi:1980pg}~\citep{Leroy:1982rg} one can define and compute the potential between $q$ and $\bar q$ external sources in the singlet and in the adjoint representation of the colour group.

Based on the theoretical approach developed in ref.~\citep{Rossi:2013qba}, in this poster we present the first few results of a set of lattice simulations carried out on a system where a $q\bar q$ pair of static colour sources is in interaction with a Yang--Mills field.

\section{External sources in the temporal gauge}
\label{sec:EXSOU}

The content of this talk is mainly based on the results of ref.~\citep{Rossi:2013qba} of which we will reproduce the relevant formulas.

The Feynman kernel in the presence of $q \bar q$-sources in the $A_0=0$ gauge reads 
\begin{eqnarray}
\hspace{-.8cm}&&K({\bf A_2},s_2,r_2;{\bf A_1},s_1,r_1;T)=\int_{{\cal G}_0}\!\!{\cal D}\mu(h) \,[U_h({\bf x}_{q})]_{s_2s_1} [U_h({\bf x}_{\bar q})]_{r_2r_1}^*\widetilde K({\bf A_2}^{U_h},{\bf A_1};T)\, ,\label{K12}\\ \hspace{-.8cm}&&\widetilde K({\bf A}_2,{\bf A}_1;T)=\int^{{\bf A}({\bf x},T_2)={\bf A}_2({\bf x})}_{{\bf A}({\bf x},T_1)={\bf A}_1({\bf x})} {\cal D}{\bf A}\exp{[-S_{YM}({\bf A}, A_0=0)]}\, ,\label{KTILDE12}
\end{eqnarray}
where ${\cal D}\mu(h)$ is the invariant Haar measure over the group, ${\cal G}_0$, of the (topologically trivial) time-independent gauge transformations that tend to the identity at spatial infinity.

The states which are the basis of the spectral decomposition 
\begin{eqnarray}
\hspace{-.7cm}&&K({\bf A}_2,s_2,r_2;{\bf A}_1,s_1,r_1;T)=\sum_{k}e^{-E_kT}\psi_{k}({\bf A}_2,s_2,r_2)\,\psi^\star_{k}({\bf A}_1,s_1,r_1)\label{KSPEC12}
\end{eqnarray}
are eigenstates of the Hamiltonian with eigenvalue $E_k$ 
\begin{eqnarray}
{\cal H}\psi_{k}({\bf A},s,r)=E_k\psi_{k}({\bf A},s,r)\, \label{H}
\end{eqnarray}
and transform covariantly under $U_w({\bf x})\in {\cal G}_0$ according to 
\begin{eqnarray}
\psi_{k}({\bf A}^{U_w},s,r)=\sum_{s',r'}\Big{[}e^{-i\lambda^a w^a({\bf x}_q)}\Big{]}_{s\,s'}\Big{[}e^{i\lambda^a w^a({\bf x}_{\bar q})}\Big{]}_{r'\,r}\psi_{k}({\bf A},s',r')\, .\label{GAUSS}
\end{eqnarray}
The scalar product in the Hilbert space of energy eigenstates with gauge transformation properties~(\ref{GAUSS}) must be defined via the Faddeev--Popov  procedure as~\citep{Rossi:1983hr}
\begin{eqnarray}
&&\hspace{2.5cm}(\psi,\phi) = \int {\cal D}\mu_F({\bf A})\sum_{s,r} \psi^*({\bf A},s,r) \phi ({\bf A},s,r) \, ,\label{SCALAR}\\
%\end{eqnarray}\begin{eqnarray}
&&{\cal D}\mu_F({\bf A})=\Delta_{F}({\bf A})\prod_{{\bf x}} \delta[F({\bf A})]\,d{\bf A}({\bf x})\, , \qquad
%\label{SCALAR}\\&&
1=\Delta_{F}({\bf A})\int_{{\cal G}_0} {\cal D}\mu(h) \delta[F({\bf A}^{U_h})]\, .\label{DELTA}
\end{eqnarray}
%\begin{eqnarray}
%&(\psi,\phi) = \int {\cal D}\mu_F({\bf A})\sum_{s,r} \psi^*({\bf A},s,r) \phi ({\bf A},s,r) \, ,\label{SCALAR}\\
%&{\cal D}\mu_F({\bf A})=\Delta_{F}({\bf A})\prod_{{\bf x}} \delta[F({\bf A})]\,d{\bf A}({\bf x})\, , \qquad
%\label{SCALAR}\\&&
%1=\Delta_{F}({\bf A})\int_{{\cal G}_0} {\cal D}\mu(h) \delta[F({\bf A}^{U_h})]\, .\label{DELTA}\end{eqnarray}
The scalar product is independent of the gauge fixing functional $F({\bf A})$. In the following numerical simulations we will choose {\small{$F({\bf A})={\nabla}{\bf A}$}}.

\section{Energy eingenstate classification}
\label{sec:ENCLAS}

The energy eigenstates in a given colour channel are classified according to irreducible representations of the colour group, which acts on the states as
\begin{equation}
{\cal U}(V) \psi({\bf A},s,r)= V_{ss'} \psi({\bf A}^V,s',r') V^\dagger_{r'r}\, .\label{TRANSF}
\end{equation} 
%${\cal U}(V)$ commutes with the Feynman kernel~(\ref{K12}), $[{\cal U}(V),K]$. 
Parametrizing the $\cal H$-eigenfunctionals in the $q\bar q$ sector as
\begin{eqnarray}
\psi({\bf A},s,r) = \phi ({\bf A}) \delta_{sr} + \phi_a ({\bf A})\lambda^a_{sr} \, ,\label{PARAM}
\end{eqnarray}
it is shown in~\citep{Rossi:2013qba} and~\citep{PROC} that the states must obey one of the following alternatives

\qquad $\bullet$ $\phi_a ({\bf 0}) =0$ and  $\phi ({\bf 0}) \neq 0$

\qquad $\bullet$  $\phi ({\bf 0}) = 0$ and $\phi_a ({\bf 0}) \neq 0$ (for some $a$) 

\qquad $\bullet$ $\phi ({\bf 0}) = \phi_a ({\bf 0}) =0$ \\
in correspondence to different types of irrep's discussed in detail in ref.~\citep{Rossi:2013qba}. As a consequence the $\cal H$-eingenstates are classified in four types according to how ''spin'' and ''orbital'' functionals transform under global colour rotations. One gets 

(1)\,\,spin singlet $\otimes$ orbital singlet states ($\phi_a ({\bf 0}) = 0$ \& ${\phi ({\bf 0}) \neq 0}$)
\begin{eqnarray}
&&\psi^{[S]}_{[S]} ({\bf A})= \phi ({\bf A}) I\, , 
\quad{\mbox{with}}\quad \phi ({\bf A}^V) = \phi ({\bf A})\in [S]_{\rm orbit}\nonumber
\end{eqnarray}
%with $\phi ({\bf 0}) \neq 0$.

(2)\,\,spin adjoint $\otimes$ orbital singlet states\, ($\phi ({\bf 0}) = 0$ \& ${\phi_a ({\bf 0}) \neq 0}$)
\begin{eqnarray}
&&\psi_{[Ad]}^{[S]}({\bf A}) = \lambda^a \phi_a ({\bf A})\, ,
%\label{coloro} 
\quad{\mbox{with}}\quad \phi_a ({\bf A}^V) = \phi_a ({\bf A})\in [S]_{\rm orbit} \nonumber
\end{eqnarray}
%with $\phi_a ({\bf 0}) \neq 0$ for some values of $a$. 

(3)\,\,spin singlet $\otimes$ orbital $[\alpha]$ states \,($\phi^{[\alpha]}_{m} ({\bf 0}) = 0$)
\begin{eqnarray}
\hspace{-.5cm}&&\psi^{[\alpha]}_{m\,[S]}({\bf A}) = \phi^{[\alpha]}_{m} ({\bf A}) I \, ,\quad{\mbox{with}}\quad \phi^{[\alpha]}_{m}({\bf A}^V) = R^{[\alpha]}_{mm'} (V) \phi^{[\alpha]}_{m'} ({\bf A})\in [\alpha]_{\rm orbit} \nonumber
\end{eqnarray}

(4)\,\,spin adjoint $\otimes$ orbital $[\beta]$ states combined in the irrep.\ $[\alpha']$ %$[\alpha]\in [\beta] \otimes [N_c^2-1]$\,\, 
($\phi_{a k} ({\bf 0}) =0$)
\begin{eqnarray}
\hspace{-.5cm}&&\psi^{[\alpha']}_{m\,[Ad]}({\bf A}) = \lambda^a \phi_{a k} ({\bf A}) \, ,\quad{\mbox{with}}\quad 
{\phi_a}_k ({\bf A}^V) = R^{[\beta]}_{kk'} (V) \phi_{a k'} ({\bf A})\in [\beta]_{\rm orbit} \nonumber
\end{eqnarray}

\section{Extracting singlet and adjoint potentials in numerical simulations}
\label{sec:ESANS}

We first of all computed on the lattice the quantity
\begin{eqnarray}
\hspace{-.8cm}&&\int {\cal D} \mu_F ({\bf A}) \!\int_{{\cal G}_0} \!\!{\cal D}\mu (h)\, [V  U_h({\bf x}_q)]_{ss} [V  U_h({\bf x}_{\bar q})]_{rr}^* {\tilde K}({\bf A}^{V U_h} ,{\bf A};\!T) = \sum_n e^{-E_n T} \chi_n (V) \, ,
\label{Totally_twisted_kernel}
\end{eqnarray}
which checks the colour content of the energy eigenstates of the theory, as a function of their colour character.

\begin{figure}[ht!]
%\hspace{-3.cm}
\centering
\includegraphics[scale=0.50,angle=0]{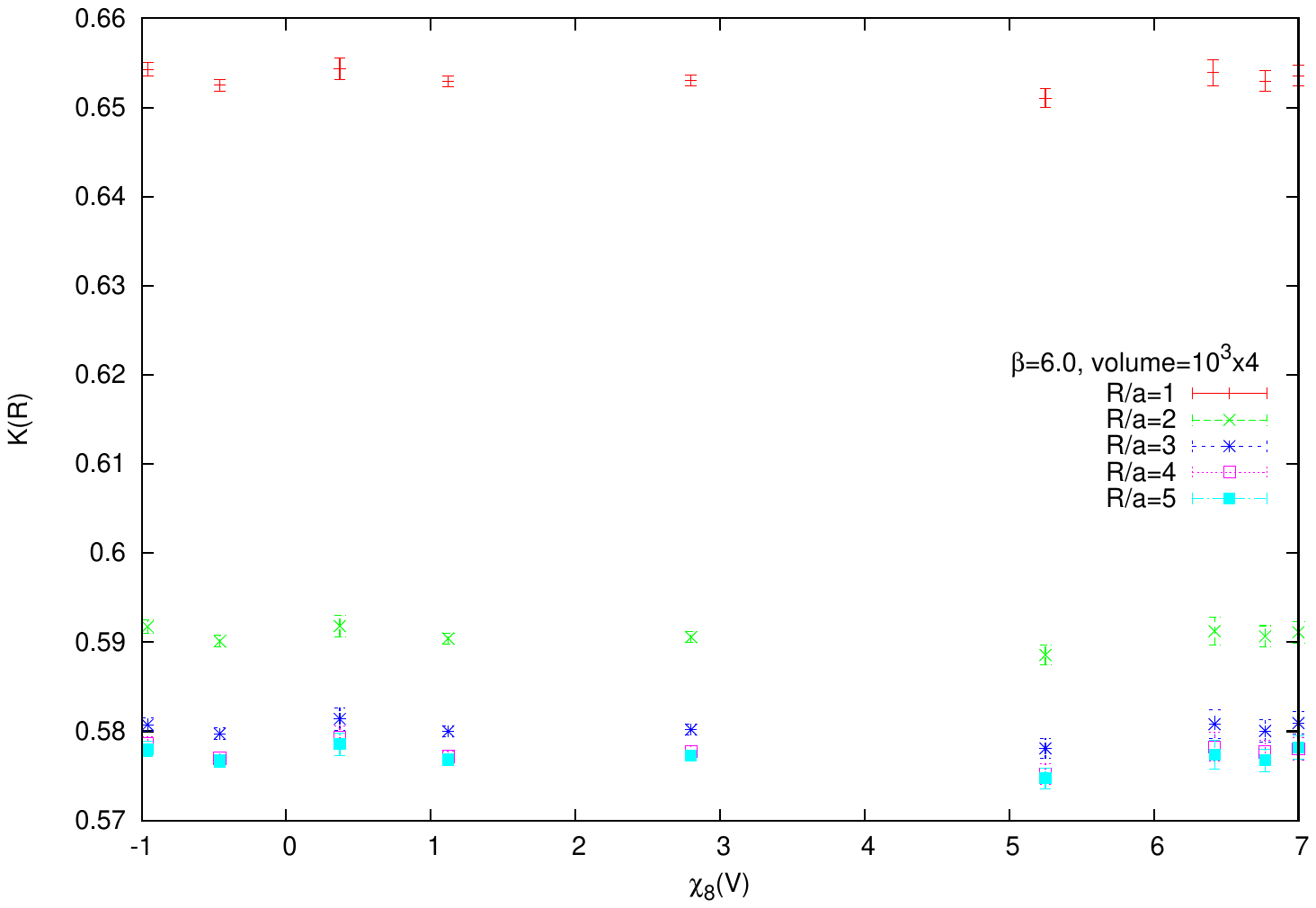}
% Global_color.pdf: 0x0 pixel, -2147483648dpi, 0.00x0.00 cm, bb=
\caption{Numerical results for eq.~(4.1)
%(\ref{Totally_twisted_kernel}) 
as a function of $\chi_8(V)$, the octet character of $V$, at various fixed distances $R=|\mathbf{x}_q - \mathbf{x}_{\bar{q}}|$.}
\label{Totally_twisted_figure}
\end{figure}

The numerical results for eq.~(\ref{Totally_twisted_kernel}), shown in Fig.~\ref{Totally_twisted_figure}, are obtained performing Monte Carlo simulations of the pure lattice gauge theory with the standard Wilson 
action on lattices of size $10^3 \times 4$ and coupling $\beta=6.0$. The quantity in eq.~(\ref{Totally_twisted_kernel}) 
is computed rotating by a matrix $V \in SU(3)$ the spatial boundary links at the final time. 
%introducing twisted boundary conditions by an $SU(3)$ matrix $V$ on the spatial links in the time direction. 
For each matrix $V$ we generated a different ensemble with a number
of lattice configurations varying from $2000$ to $10000$. The statistical errors are estimated using a jackknife algorithm. 

%\newpage 
%where $R$ denotes the space distance of the two sources.

The $V$ independence of the plot in Fig.~\ref{Totally_twisted_figure} seems to suggest that only global colour singlets contribute to the kernel in eq.~(\ref{KSPEC12}). 
This fact may be due to either colour confinement,  
%This fact may be due to two possible reasons. The first possibility could be due to colour confinement, 
which would imply that only global colour singlet states survive in the theory, or 
%. Another possibility, 
as discussed in refs.~\citep{Rossi:2013qba} and~\citep{PROC}, that in the numerical simulations the integration over the group of topologically trivial gauge transformations, is effectively extended to the group $\overline{\cal G}$ also including global colour rotations. This colour averaging would have the effect of 
%of ``all'' the possible gauge transformations, thus 
washing out the contribution of the colour non-invariant states. Further numerical investigation is needed to discriminate between these two possibilities.

A step in this direction was to compute on the lattice the quantity 
\begin{eqnarray}
&&\overline{\mbox{K}}{\mbox{(R;V)}} \equiv \int {\cal D} \mu_F ({\bf A}) \int_{\overline{\cal G}} {\cal D}\overline\mu (h)\, [\overline U_h({\bf x}_q)]_{ss} [\overline U_h({\bf x}_{\bar q})]_{rr}^* {\tilde K}({\bf A}^{V^\dagger\overline U_h} ,{\bf A};\!T) \nonumber \, .
\label{Partially_twisted_kernel}
\end{eqnarray}
for which one can prove~\citep{Rossi:2013qba} the expansion formula 
\begin{eqnarray}
\overline{\mbox{K}}{\mbox{(R;V)}} = \sum_n e^{-E_n T} \chi^{\rm orb}_n (V) \nonumber \, ,
\end{eqnarray}
where $\chi^{\rm orb}_n (V)$ are the characters of the colour group representations to which the ``orbital'' (i.e.\ gluon) wave functions contributing to eq.~(\ref{KSPEC12}) belong.

\begin{figure}[ht]
 \centering
 \includegraphics[scale=0.50,angle=0]{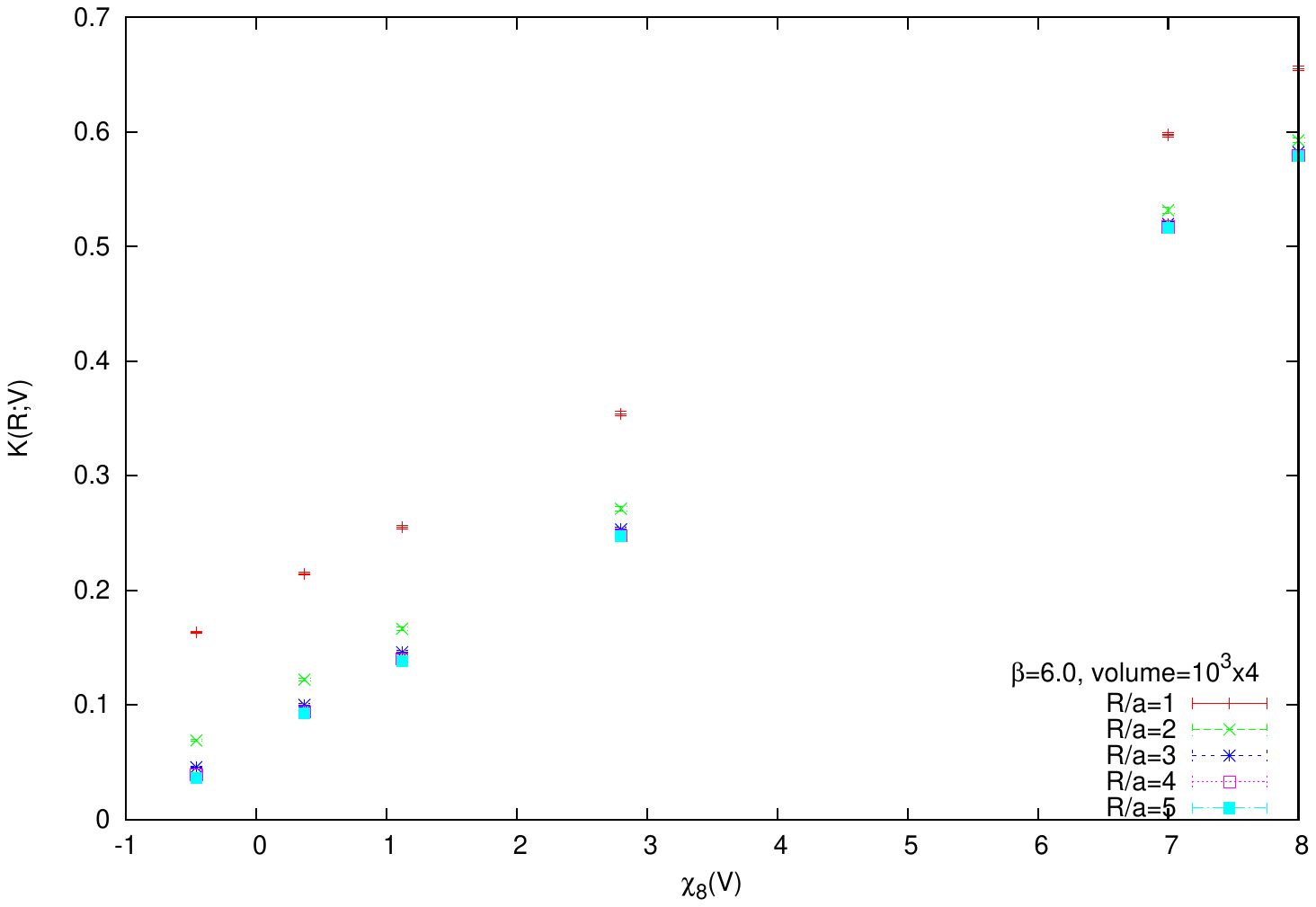}
  %Former{Orbital_color.pdf} 
 %\vspace{-3.5cm}
 % Global_color.pdf: 0x0 pixel, -2147483648dpi, 0.00x0.00 cm, bb=
\caption{Numerical results for eq.~(4.2)
%(\ref{Partially_twisted_kernel}) 
as a function of $\chi_8(V)$ at various distance $R=|\mathbf{x}_q - \mathbf{x}_{\bar{q}}|$.}
%{\small{\begin{eqnarray} &&\int {\cal D} \mu ({\bf A}) \int_{{\cal G}_0} {\cal D} \Omega \, (\Omega)_{r r} ({\bf x}) (\Omega)_{s s}^* ({\bf y}) {\tilde K}({\bf A}^{V \Omega} ,{\bf A};T) = \nn\\&&= \sum_n e^{-E_n T} \int \delta \mu ({\bf A}) \psi_n({\bf A}^V,r,s) \psi^*_n ({\bf A},r,s) = \nonumber \\&&=\sum_n e^{-E_n T} \chi_n^{(orb)} (V) \end{eqnarray}}}
\label{Partially_twisted_figure}
\end{figure}
% \vspace{3.cm}

  \newpage
The numerical results we found for $\overline{{K}}{{(R;V)}}$ (eq.~(\ref{Partially_twisted_kernel})) are shown in Fig.~\ref{Partially_twisted_figure}.
% Since  we need a gauge fixing procedure in the evaluation of the 
% quantity in eq. (\ref{Partially_twisted_kernel}), numerically expensive, we use only a subset of the twisted configurations used to
% evaluate eq. (\ref{Totally_twisted_kernel}).
Since a numerically expensive gauge fixing procedure is needed for the evaluation of this quantity, we used only a subset of the configurations used to evaluate eq.~(\ref{Totally_twisted_kernel}). Fig.~\ref{Partially_twisted_figure} seems to indicate that only ``orbital'' octets and singlets contribute to eq.~(\ref{KSPEC12}). Putting this information together with the previous observation that only global colour singlets appear in the theory, we are led to the conclusion that the singlet ``orbital'' wave function must be associated to the singlet ``spin'' function, while the octet ``orbital'' wave function must be associated to the octet ''spin'' function. 
%MODIFICA 
%This conclusion is checked in the diagram
%MODIFICA
%\newpage

\begin{figure}[ht]
 \centering
 \includegraphics[scale=0.50,angle=0]{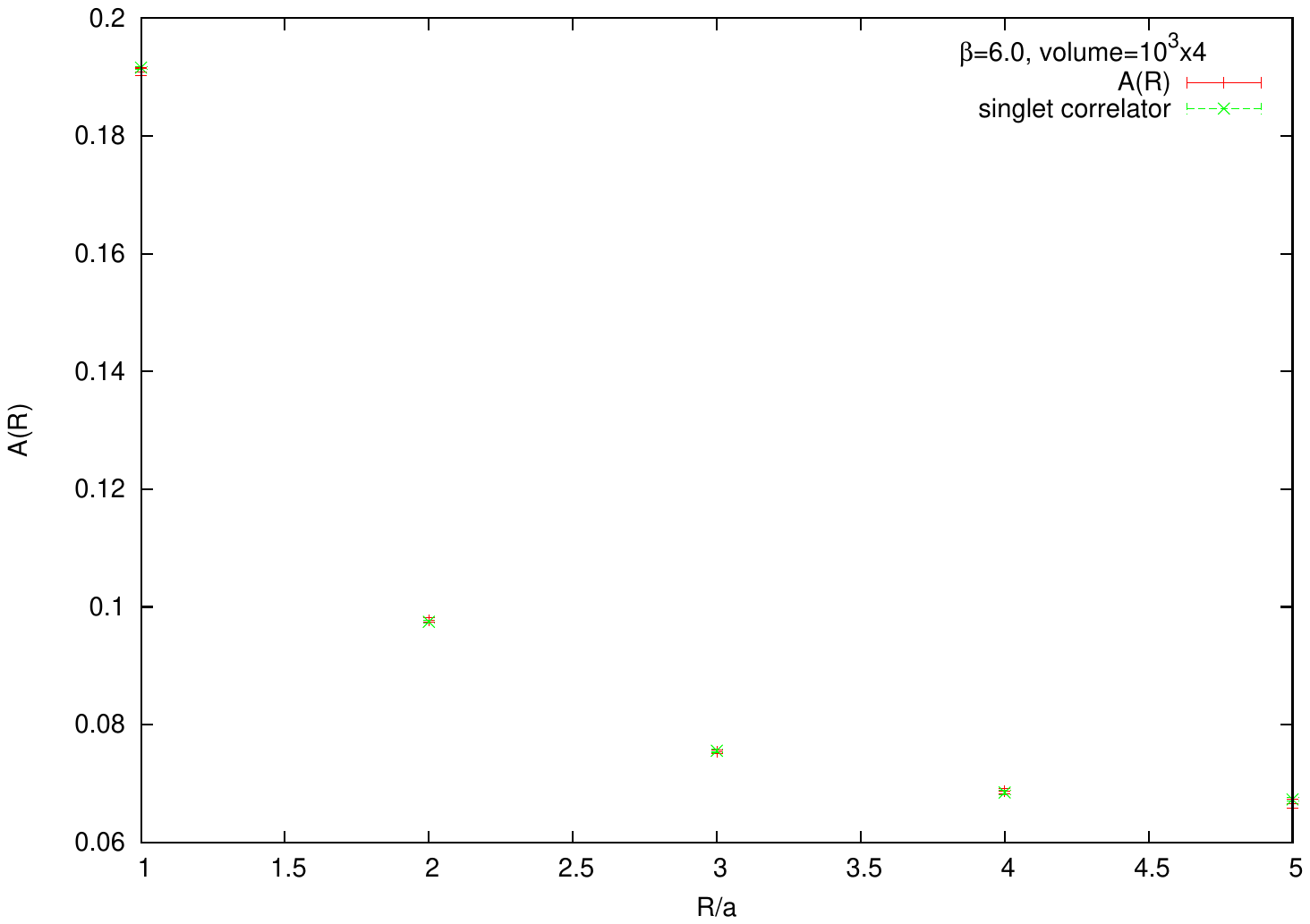}
%Former{SINGSING_COLORATO.pdf}
% Global_color.pdf: 0x0 pixel, -2147483648dpi, 0.00x0.00 cm, bb=
\caption{Comparison between the fit parameter $A(R)$ and the singlet correlator at $\beta=6.0$ and volume $10^3 \times 4$.}\label{fig:tre}
%\vspace{-3.5cm}
 \label{SingletSinglet}
 \end{figure}

\begin{figure}[ht!]
 \centering
 \includegraphics[scale=0.50,angle=0]{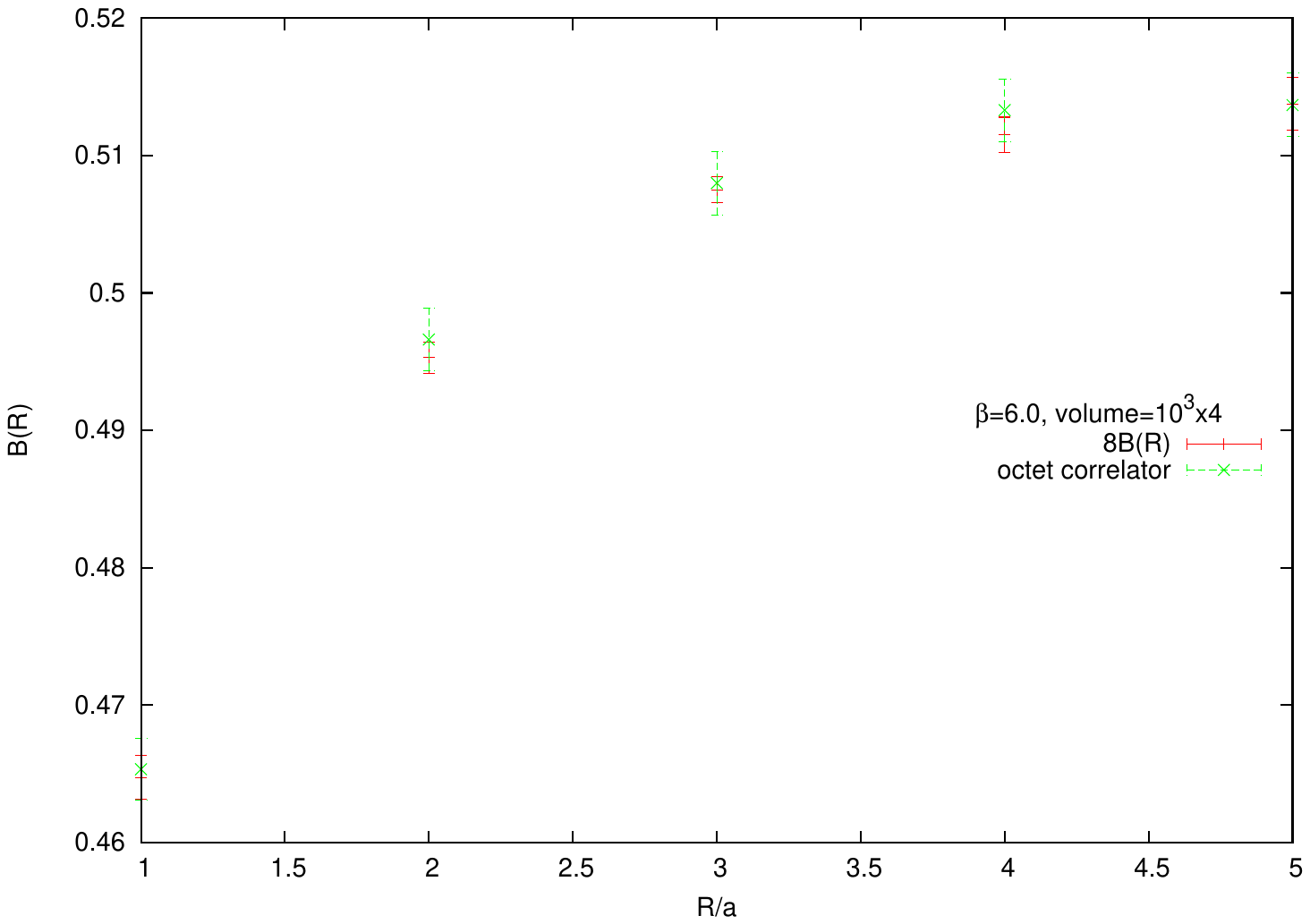}
 %\vspace{-3.cm}
%Former{OCTOCT_COLORATO.pdf}
% Global_color.pdf: 0x0 pixel, -2147483648dpi, 0.00x0.00 cm, bb=
\caption{Comparison between the fit parameter $B(R)$ and the octet correlator at $\beta=6.0$ and volume $10^3 \times 4$.}\label{fig:quattro}
\label{OctetOctet}
\end{figure}

These conclusions are confirmed by the data shown in Figs.~\ref{SingletSinglet} and~\ref{OctetOctet}.
If we parametrize the results of the quantity in eq.~(\ref{Partially_twisted_kernel}) in the form 
\begin{equation}
 \bar{K}(R;V)=A(R) + \chi_8(V)B(R),
\end{equation}
where 
\begin{equation}
 A(R)=\sum_n e^{-E^{[S]}_n(R)T}\label{AR}
\end{equation}
and
\begin{equation}
 B(R)=\sum_n e^{-E^{[Adj]}_n(R)T} \, ,\label{BR}
\end{equation}
one can extract the values of the parameters $A$ and $B$ from a linear fit to the numerical data %for eq.~(\ref{Partially_twisted_kernel}) 
in Fig.~\ref{Partially_twisted_figure}.

 \newpage
In Fig.~\ref{SingletSinglet} we compare the potential energy of the singlet ``spin'' contribution extracted from the singlet character dependence (eq.~(\ref{AR}) and Fig.~\ref{Partially_twisted_figure}) and the same quantity computed by means of the gauge fixing procedure of the total trace of the Feynman kernel, i.e.\ from the singlet-projection 
\begin{eqnarray}
\sum_{s_2r_2s_1r_1}\!\!\frac{1}{N_c}{\delta_{r_2s_2}\delta_{s_1r_1}} \int {\cal D}\mu_F({\bf A}) K({\bf A},s_2,r_2;{\bf A},s_1,r_1;T) \, .
\end{eqnarray}
This quantity is computed on configurations generated with the same Monte Carlo algorithm and lattice parameters as before, but imposing periodic boundary conditions also in the time direction.
% We choose to fix the Coulomb gauge on the slice at $t=0$.
The plot in Fig.~\ref{SingletSinglet} clearly shows that the two ways of evaluating $A(R)$ lead to perfectly consistent results.

The same procedure applied to the octet leads to the plot in Fig.~\ref{OctetOctet}.
The lattice quantity to be compared with $B(R)$ (eq.~(\ref{BR})) is evaluated by computing the gauge fixed octet-projection 
\begin{equation}
 2\sum_{s_2r_2s_1r_1a}\!\! \lambda^a_{r_2s_2}\lambda^a_{s_1r_1} \int {\cal D}\mu_F({\bf A}) K({\bf A},s_2,r_2;{\bf A},s_1,r_1;T) \, .
\end{equation}

%\newpage

%\newpage
\noindent Fig.~\ref{OctetOctet} again shows perfect agreement between the two ways of computing $B(R)$.

\section{Conclusions}
\label{sec:CONCLO}

In this talk we have presented some preliminary results of numerical simulations with the purpose to study, in a non-perturbative context, the results of ref.~\citep{Rossi:2013qba}.

\acknowledgments 
We thank G.C. Rossi for many useful discussions.

\end{document}